# Dispersion-compensated meta-holograms based on detour phase

Mohammadreza Khorasaninejad[1,*], Antonio Ambrosio[1,2,*], Pritpal Kanhaiya[1,3] and Federico Capasso[1]

[1]*Harvard John A. Paulson School of Engineering and Applied Sciences, Harvard University, Cambridge, Massachusetts 02138, USA*

[2]*CNR-SPIN U.O.S. Napoli, Dipartimento di Fisica, Università di Napoli Federico II, Complesso Universitario di Monte Sant'Angelo, Via Cintia, 80126 - Napoli, Italy*

[3]*University of Waterloo, Waterloo, ON N2L 3G1, Canada*

*These authors contributed equally to this work

*Corresponding author: capasso@seas.harvard.edu*

Subwavelength structured surfaces, known as metasurfaces, hold promise for future compact and optically thin devices with versatile functionalities. Here, by revisiting the concept of *detour phase* at the basis of the first computer generated holograms we have designed high-performance transmissive dielectric meta-holograms with new functionalities. In one class of such devices wavelength-independent phase masks have been generated by compensating the inherent dispersion of the detour phase with that of the subwavelength structures. This leads to broadband operation from the visible to the near infrared with efficiency as high as 75% in the 1.0 to 1.4 μm range. We have also shown that we can fully control the effective focal length of the imaging optical system by incorporating a lens-like function in the meta-hologram. In a second class of devices we have incorporated in the phase map a geometric phase to achieve for the first time chiral imaging: the projection of different images depends on the handedness of the reference beam. The compactness, lightweight and ability to produce images even at large angles of our devices have significant potential for important emerging applications such as wearable optics.

INTRODUCTION

Holography, the branch of optics that uses the *whole* (*ὅλος* in Greek) information, amplitude and phase, of light represented a major step forward in optical science in the second half of the last century [1]. Although initially an interference pattern had to be recorded in photographic plates, the principle of holography has since greatly expanded visualization possibilities [2-3] and applications [4-5], in particular through the widespread use of computer-generated holograms [6]. Holographic principles have been applied to overcome experimental challenges where other approaches have failed. This has been the case, for instance, with imaging through diffusive media [7] and the realization of portable 3D projectors [4,8]. Nowadays, the enormous progress in nano-fabrication techniques may revolutionize holography just as the invention of the laser did many years ago. Subwavelength structured optical elements (gratings, etc.) [9], and more generally metamaterials [10] and metasurfaces [11-30], with their ability to control the phase, amplitude, and polarization over subwavelength scales without requiring volumetric propagation, are opening new frontiers in holographic and optical devices such as wearable displays. However, to the best of our knowledge, in all of these implementations, the wavelength response is highly dispersive and highly efficient operation has been limited to reflection-mode configurations.

In this work, we show how basic holographic principles can be revisited to implement new capabilities of wavefront molding with planar subwavelength dielectric optical elements. Our holographic devices are transmissive, broadband and phase distortion-free from the near infrared (NIR) to the visible with high polarization sensitivity allowing for a wide range of functionalities depending on the design. The basic element of our holograms (Fig. 1) is an "effective aperture" designed to diffract light with high efficiency across a broad range of wavelengths into the (±1) diffraction order; thus it is functionally equivalent to a broadband blazed grating. The phase profile of our meta-holograms is then generated by displacing the effective apertures with respect to each other in such a way to create the desired interference pattern. The latter is the detour phase concept at the basis of binary holograms. A complex computer generated hologram representing the logo of the International Year of Light has been imaged with high efficiency, up to 75% in the NIR. By further incorporating a lens in the hologram, the effective focal length of the imaging system has been shifted by a controlled amount. Polarization selective holograms have been widely investigated [20-22]; here we have demonstrated a chiral holographic-plate that creates different images depending on the handedness of the incident light.

DESIGN OF DISPERSION-LESS META-HOLOGRAMS

It goes back to Lord Rayleigh and later Michelson the observation of how periodic errors in grating fabrication affect the interference pattern by modulating the wavefront [31]. This concept of wavefront shaping through displacement of adjacent elements, known as detour phase, has been later perfected and used in the first computer generated holograms. Detour phase is also the design principle of our devices [2, 32, 33] and it is the core concept behind binary holograms where the amplitude and phase of the optical field are imposed by an array of apertures on an opaque screen. In a typical binary hologram, the dimension of each aperture sets the amount of light passing through it. Instead, the phase shift of the light diffracted from two adjacent apertures along a given direction ($\theta$) (Fig. 1(a)) is controlled by adjusting the distance between them. The light wavelets from the two apertures are in phase if the distance $D$ between them is $n\lambda/sin(\theta)$, where $n$ is an integer and $\lambda$ is the wavelength. For any other distance the wavelets from the two apertures will be phase shifted relative to each other by the amount:

$$\Delta\varphi = \frac{2\pi D}{\lambda}\sin(\theta) \qquad (1)$$

Note that the wavelength dependence of this phase shift can be suppressed if diffraction from the apertures is designed in such a way as to compensate the intrinsic dispersion. From Eq. (1) one can see that this dispersion-less condition can be achieved if each aperture is replaced by a subwavelength structured element (meta-element) with an engineered dispersion similar to that of a grating, i.e. $\sin(\theta) \sim \lambda$.

Current lithography techniques allow surface structuring with subwavelength resolution. In our devices the apertures are replaced by effective apertures (pixels) with a polarization functionality provided by the specifically incorporated meta-element. For the first study, the meta-element consists of three dielectric ridge waveguides (DRWs) made of amorphous silicon (a-Si), on a glass substrate (Fig. S1 of Supplementary Materials (SM)). A diffraction condition where the majority of the transmitted light is funneled into the first orders (±1) is achieved by adjusting the DRWs design parameters (width, height, and separation) and the lateral dimension $\Lambda$ of the meta-element (Fig. S2 of SM). In our design, each pixel of dimension $2\Lambda$ consists of two of such meta-elements, i.e. six amorphous silicon ridges (Fig. 1(b)). Figure 1(c) shows that the transmitted power from each pixel is nearly completely split between the ±1 orders in a broad wavelength range (1100 nm - 1800 nm), while unwanted orders are suppressed. The dispersive response of the pixel is designed with finite difference time domain (FDTD) simulations to closely satisfy the relation $\theta = sin^{-1}(\lambda/\Lambda)$ (Fig. 2(a)), so as to achieve a wavelength-independent phase shift $\Delta\varphi$, as previously discussed.

If we now design a hologram out of these effective apertures, the phase shift of the light diffracted by each aperture along the direction $\theta$, defined relative to the light from a reference aperture propagating along the same direction $\theta$, is still given by Eq. (1). Substituting $\theta = sin^{-1}(\lambda/\Lambda)$ into the Eq. (1) then yields the dispersion-less detour phase:

$$\Delta\varphi(x_m, y_m) = \frac{2\pi D(x_m, y_m)}{\Lambda} \qquad (2)$$

where $(x_m, y_m)$ are the coordinates of the pixels' centers and the index $m$ spans over the total number of pixels.

Note that, although the description of our pixels as effective apertures is useful, in our device there is no opaque screen or apertures but a transparent glass substrate whose surface is structured in pixels displaced with respect to each other in order to obtain the required phase map. The desired phase distribution that will generate the intensity pattern of interest when illuminated by the reference beam is then computed by means of the widely used Gerchberg-Saxton phase-retrieval algorithm [34] (see SM). The computed phase map is then converted into a spatial distribution of displacements $D(x_m, y_m)$ that defines the equivalent of a binary phase-only hologram. In our design, the light diffracted by two adjacent pixels is in phase if the pixels' centers are spaced by $2\Lambda$, i.e. the two effective apertures are touching (Fig. 1(b)). The desired phase modulation is then achieved by displacing each pixel with respect to the in-phase condition. This design represents the highest possible pixel density. We chose this configuration in order to maximize the total cross-section of the device to the illuminating light, minimizing the spacing between adjacent pixels and the light that passes through non-diffracted. This configuration results in the physical overlap of a certain amount of effective apertures. In the design process, we removed the DRWs that physically overlap (less than 5% of the total number) while we kept all the six ridges when the pixel overlap only affects the area without DRWs. An example of this condition is highlighted in the inset of Fig. 2(c). The pixels containing blue and red DRWs partially overlap by design. Such partial overlap only affects the amount of light along *θ* from those pixels and results into a possible decrease of the contrast of the image in the Fourier plane (random amplitude modulation). However, the high-quality of the images obtained proves that this effect is negligible in our devices. Furthermore, we note that this type of phase modulation allows 0-2π continuous phase variation and has a high tolerance to the fabrication process, which in the present case is electron beam lithography (EBL). Our EBL (ELS-F 125) has an ultra-high beam positioning resolution (0.01 nm) that results in minimum phase mismatch [2] in the detour phase (Fig. S3 of SM). Design tolerances are more demanding for those approaches where the phase modulation is achieved by changing the shape of subwavelength resonators or scatterers.

RESULTS AND DISCUSSION

As an example of our device functionality, a far-field intensity distribution corresponding to the 2015 International Year of Light (IYL) logo is designed [35]. This device consists of 256×256 pixels as shown in Fig. 2(b). Figure 2(c) shows the scanning electron micrograph (SEM) of the fabricated device (see Fig. S4 of SM for the optical image of the device). A sketch of the experimental setup for image acquisition is shown in Fig. 2(d). A thin lens is used to obtain an image in the Fourier plane where the camera sensor is placed. Figure 3(a)-(d) show images of the intensity distribution generated by the hologram in the Fourier plane under NIR illumination. By design, the hologram provides the same image quality for all the wavelengths in the range of *λ = 1100-1800 nm*, which is limited by our measurement setup (camera sensitivity and laser source range). This feature results from the dispersion-less phase realization approach (Eq. (2)). We note that the dimension of the images in the reconstruction plane varies with wavelength according to the well-known relation ($N_i \times \lambda \times f$ )/$L_i$ where $f$ is the focal length of the lens while $L_i$ and $N_i$ are the device dimensions and number of pixels along the *x*- and *y*-axes, respectively [2].

The measured efficiency as a function of wavelength is reported in Fig. 3(e). The efficiency is measured as the ratio of the total intensity of the first diffraction orders to the incident intensity. An efficiency as high as 75 % is achieved. Note that this value is close to the theoretical value of 81% of a binary phase grating optimized for a specific wavelength [3]. Although our device is designed as a phase-only hologram, the use of subwavelength diffractive elements makes it possible to diffract light with high efficiency in the first orders only. The wavelength-dependence of the efficiency can be interpreted in terms of the angular distribution of the diffracted power. In fact, from the simulation in Fig. 1(c) it is evident that for wavelengths between 1150 nm and 1550 nm most of the light goes only to the first orders with a small (less than 10%) contribution to the zero-order. However, as the wavelength is reduced, the angular response of the first orders narrows, resulting in a peak efficiency around 1250 nm. Figure 3(h)-(k) show that the hologram maintains its functionality even for visible light (see also Fig. S5 and Fig. S6) based on its dispersion-less design, although the transmittance of the device drops towards shorter wavelengths due to absorption from the silicon ridges.

The interaction of the DRWs with the incident light is highly polarization dependent due to the DRWs' deep-subwavelength width and asymmetric cross-section (the length is much larger than the width) [24]. In other words, only light linearly polarized along the length of the DRWs (*y*-axis) is efficiently diffracted while light polarized along the width (*x*-axis) is transmitted nearly non-diffracted (Fig. S7). Due to the polarization sensitivity of each pixel, our device is characterized by a high extinction ratio (ER) between orthogonal polarizations, within a broad wavelength range (Fig. 3(e)). ER is defined as the ratio of normalized intensities in the images for two different polarizations, (along the *y-* and *x*-axes for the IYL hologram and circularly left- and right-polarized for the chiral hologram, to be discussed later in the paper).

This straightforward and reliable phase shift realization provides an opportunity for designing multifunctional devices. We further tested the capability of this concept by incorporating a Fresnel lens-like function to the phase map that corresponds to the 2015 IYL logo. This moves the reconstruction plane two centimeters forward in the light propagation direction. Figure 3(f)-(g) show the image at the previous focal position compared to the image in the new focal position, respectively. Image blurring and correct focusing are evident. It is worth noting that light focusing by the Fresnel hologram is occurring along the diffraction direction (θ= 30° at $\lambda$=1350 nm). We note that projection at large angles (for instance θ ~ 42° at λ=1800 nm) is a long-standing challenge (known as the shadow effect) which is overcome by our design. This opens the possibility to design flat and compact optical components for imaging at a wide range of angles.

As further proof of the versatility of our approach, we designed a chiral hologram whose functionality depends on the handedness of the reference beam. For this task, the meta-element consists of nanofins similar to those in reference [25, 26]. When circularly polarized light passes through these structures, it is diffracted along a principal direction ($\theta$) according to its handedness. Changing the light handedness results in switching of the direction from $\theta$ to $-\theta$ analogously to other chiral subwavelength structured surfaces [26, 29, 36]. According to what has been discussed so far, a phased array of such subwavelength structured pixels would result in the appearance of the designed intensity distribution along a specific direction only for circularly polarized light with the proper handedness. However, in the fabricated device, we chose to divide each pixel in two parts (along the *y*-axis direction) and have half of it couple to one handedness and the other half to the opposite handedness (Fig. 4(a)-(b)). Note that the rotated nanofins in our design introduce a geometrical phase similar to the rotated apertures of Ref. [37] known as the Berry-Pancharatman phase. In this way, the image displayed in the field of view of the optical system depends on the light handedness. Figure 4(c)-(e) show that a light intensity distribution corresponding to the letter *"R"* appears under right-circularly polarized illumination while it changes into the letter *"L"* for left-circular polarization. In this case a quarter wave-plate is used in the setup of Fig. 2(d) to generate the circular polarization. For linear polarization both letters appear. In addition, this device demonstrates high values of absolute efficiency and ER as well as a broadband functionality (Fig. S8 and Fig. S9 of SM).

In conclusion, we used meta-elements to build phased pixels in flat and compact dielectric holograms with a broadband response. Depending on the subwavelength structured building block, different responses to light polarization states can be encoded for scalable polarimetric devices. Using dielectric materials instead of metals allows one to work in a transmission scheme with a transparent substrate while minimizing the optical losses. Furthermore, lens-like optical elements working off-axis can be implemented for wearable devices where lightness, compactness and image quality are mandatory. In fact, although dynamic diffractive optical components, such as spatial light modulators (SLMs), have typically a broadband response, they have a large footprint with respect to our devices. Here we have shown that true optical functionality such as imaging at an angle can be achieved with a thin, small, lightweight and efficient diffractive element fabricated on a transparent substrate that can be easily integrated in designing near-to-eye displays and wearable optical systems. In addition, a new hologram with chiral imaging functionality has been demonstrated. Alternative fabrication methods such as deep ultraviolet lithography and nano-imprinting can facilitate the mass production of our proposed devices.

METHODS

**Evaluation of efficiency and extinction ratio.**

The efficiency is defined as the ratio of the total intensity of the first diffraction orders to the incident intensity. For the efficiency measurements the intensity was measured by substituting the InGaAs camera (Fig. 2(d)) with a NIR photodetector (Thorlabs DET10D). The incident intensity was instead measured as light passing through an aperture (aluminum on a glass) with the same size of our holographic device. Extinction Ratio (ER) is defined as the ratio of normalized intensities in the images for two different polarizations, along the y- and x-axes for the IYL hologram and circularly left- and right-polarized for the chiral hologram.

**Acknowledgments:** Fabrication work was carried out in the Harvard Center for Nanoscale Systems, which is supported by the NSF. We thank E. Hu for the supercontinuum laser (NKT "SuperK").

**Funding:** This work was supported in part by the Air Force Office of Scientific Research (MURI, grant# FA9550-14-1-0389), Google Inc. and Thorlabs Inc.

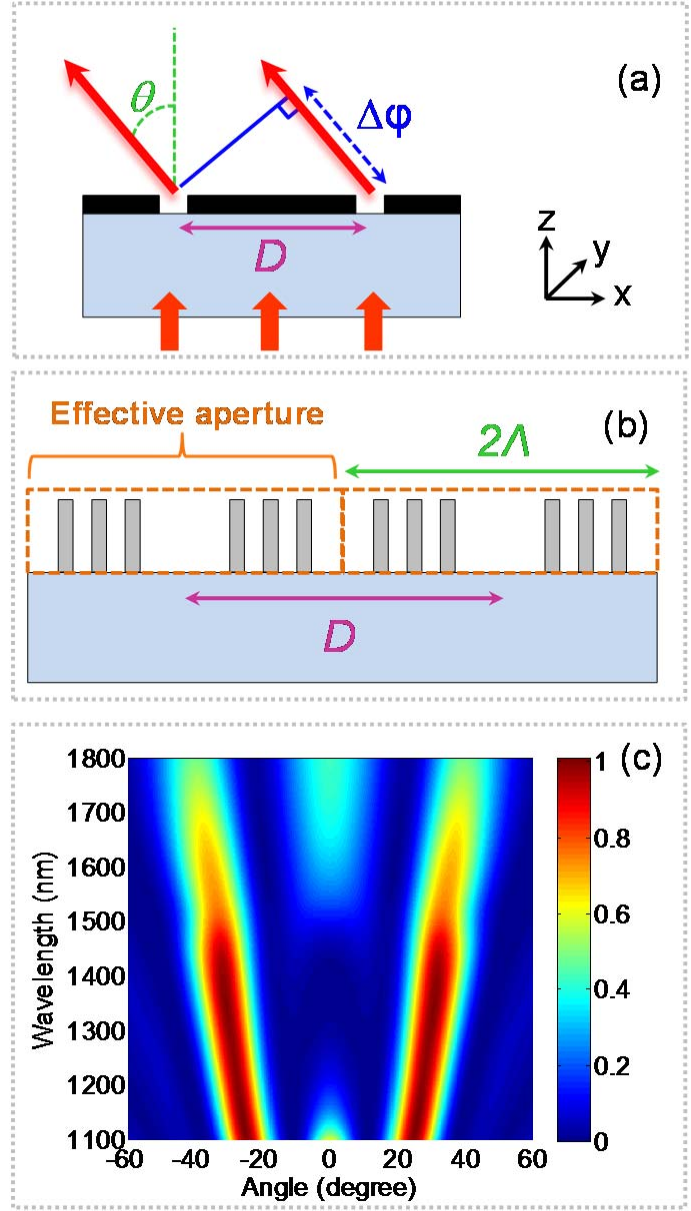

**Fig. 1. Effective aperture made of subwavelength structured dielectric.** (a) Schematic diagram of two apertures separated by a center-to-center distance $D$ on a glass substrate. The phase shift associated with light propagating from the two apertures along the direction $\theta$ is denoted as $\Delta\varphi$. This phase difference is related to $D$ and the wavelength by Eq. (1). (b) Schematic diagram of cross-section of two generic pixels ($2\Lambda=5.4$ μm) of the hologram. Each pixel functions as an effective aperture consisting of six dielectric ridge waveguides (DRWs) with subwavelength spacing ($S=380$ nm), width ($W=120$ nm) and height ($H=400$ nm). (c) Far-field (Real $(E_y)^2$) response of the meta-element when the incident light is polarized along the $y$-axis. This represents a 2D finite difference time domain simulation, where the DRWs are infinitely long along the $y$-axis while the meta-element extension along

*x*-axis is *2Λ=5.4 μm*. Engineering the dispersive response of the DRWs results in highly-directional diffraction in which the majority of the transmitted light is funneled into the first orders while other diffraction orders are suppressed.

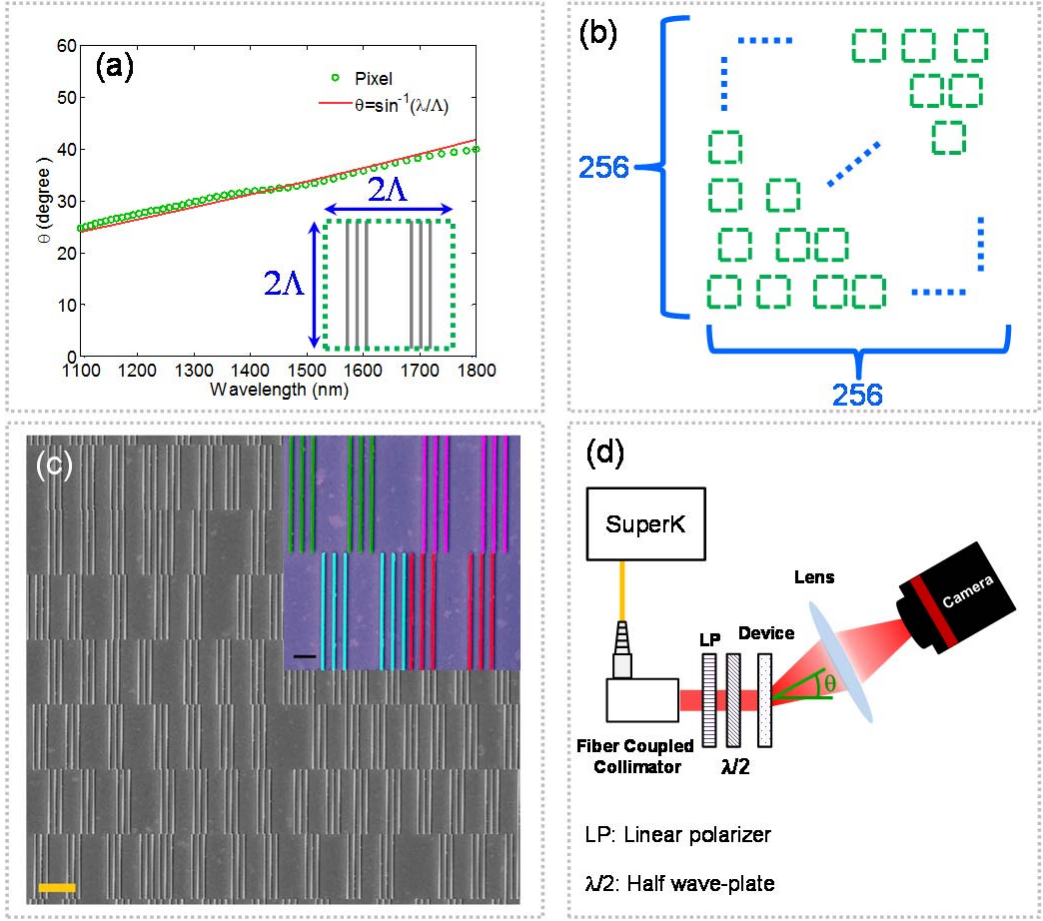

**Fig. 2 (a). Dispersion engineering and hologram design**. Diffraction angle of a pixel as a function of wavelength. The inset shows top-view of the hologram pixel with size 5.4 μm ×5.4 μm (2$\Lambda$=5.4 μm). The deflection angle is calculated from the far-field response using finite difference time domain simulations. The modeled deflection angle closely follows the target angular dispersion $\theta=sin^{-1}(\lambda/\Lambda)$ in order to cancel the wavelength dependence of the detour phase. (b) Schematic diagram showing the 256×256 pixels arrangement in the hologram. The required phase map is achieved through the displacement of each pixel. (c) Scanning electron micrograph (SEM) of the device. The scale bar is 1 μm. The inset shows a false colored SEM image in which four pixels are highlighted with different colors. (d) Sketch of the experimental setup: the laser beam from a fiber coupled Supercontinuum laser is collimated by means of a fiber collimator. The polarization state of the laser beam is controlled by a linear polarizer (LP) followed by a half wave-plate ($\lambda$/2) before illuminating the hologram (device). An image of the light distribution in the Fourier plane is obtained by means of a lens and an InGaAs camera aligned along the direction of the first diffraction order. The lens is at a focal length distance from both the device and the camera. For measurements in the visible the InGaAs camera is substituted by a color camera.

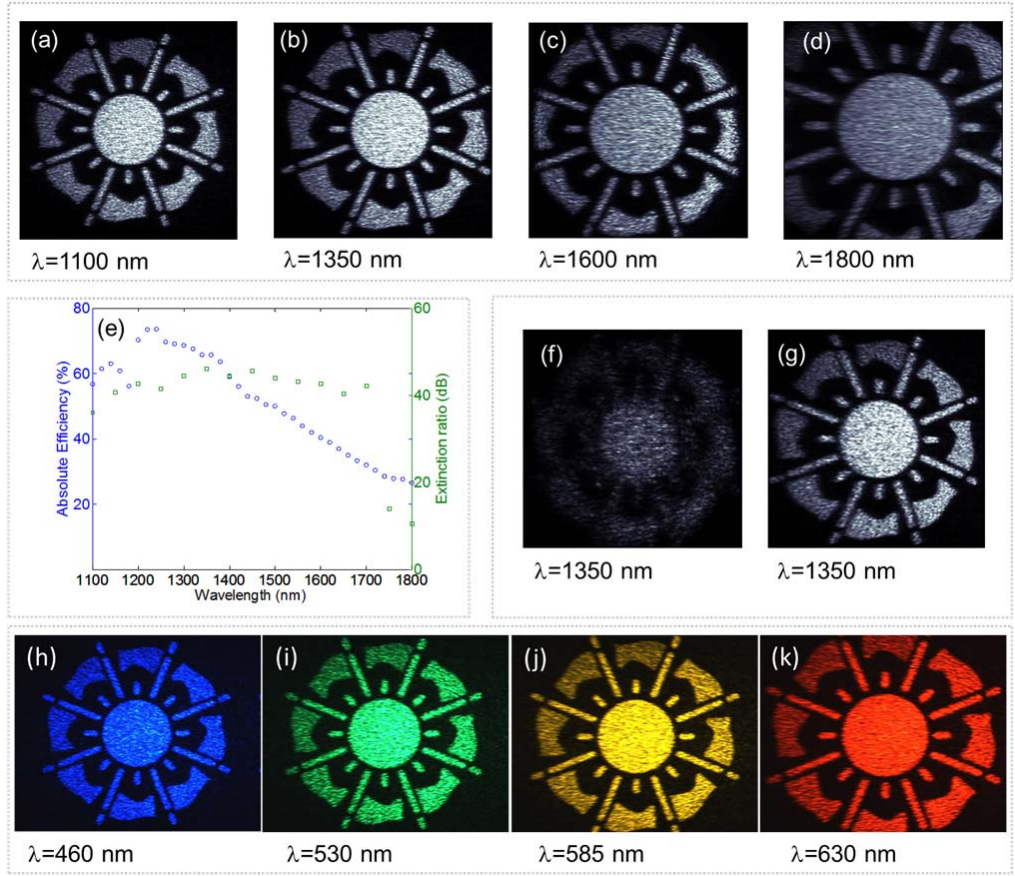

**Fig. 3. Broadband phase-distortion-free hologram.** (a)-(d) Images generated when the hologram is illuminated with near infrared light. (e) Absolute efficiency and extinction ratio (ER) as a function of wavelength. The drop in ER is due to the significant drop in the efficiency of the imaging camera at longer wavelengths. (f)-(g) Images corresponding to a hologram in which the International Year of Light logo phase distribution is added to that of a Fresnel lens with a total shift of the reconstruction plane of 2 cm. (f) Image captured under the same measurement conditions of Fig. 3 (a)-(d). This image is blurry since the Fresnel lens phase profile encoded in the hologram moves the image plane two centimeters forward along the propagation direction. (g) The same image appears correctly in focus when the camera is moved 2 cm along the propagation direction. (h)-(k) Images generated by the hologram in the visible range. These images were captured by a color CCD camera.

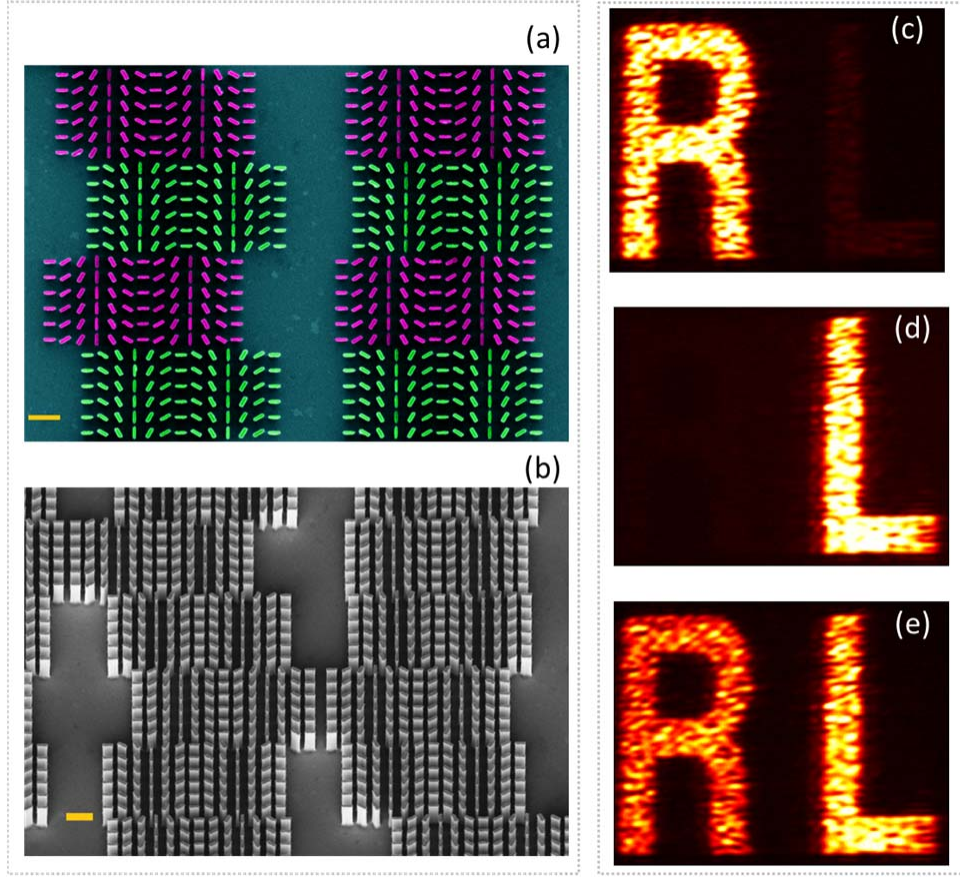

**Fig. 4. Chiral hologram.** (a) False colored scanning electron micrograph (SEM) of four pixels of the hologram. Each pixel consists of two parts: in purple, those that impart the required phase map for letter "L" and in green, those for the phase map for letter "R". Nanofins have width *W=85 nm*, length *L= 350 nm*, height *H=1000 nm*, and center-to-center distance of *500 nm*. The scale bar is 1 μm. (b) Tilted-view: SEM image of the hologram. The scale bar is 1 μm. (c)-(e) Images in the +1 diffraction order (false colored) generated by the chiral hologram under different incident polarizations at *λ*=1350 nm. Chiral hologram illuminated by (c) right-circularly (d) left-circularly and (e) linearly polarized light resulting in the appearance of the letters "*R*", "*L*" and "*RL*", respectively.